\documentclass[pre,superscriptaddress,twocolumn]{revtex4-1}

\usepackage{amsmath}
\usepackage{epsfig}
\usepackage{array}
\usepackage{times}
\usepackage{setspace}
\usepackage{mathrsfs}
\usepackage{amssymb}
\usepackage{color}

\begin{document}

\title{Heat-machine control  by quantum-state preparation: from quantum engines to refrigerators}

\author{D. Gelbwaser-Klimovsky}
\affiliation{Weizmann Institute of Science, 76100
Rehovot, Israel}
\author{G. Kurizki}
\affiliation{Weizmann Institute of Science, 76100
Rehovot, Israel}

\begin{abstract}
We explore the dependence of the performance bounds of heat engines and refrigerators on the \textit{initial quantum state} and  the subsequent evolution of their  piston, modeled by a quantized harmonic oscillator.   Our goal is to provide a fully quantized treatment of \textit{self-contained (autonomous)} heat machines, as opposed to their prevailing semiclassical description that consists of a quantum system alternately coupled to a hot or a cold  heat bath, and \textit{ parametrically} driven by a classical time-dependent piston or field.
Here by contrast, there is no external time-dependent driving.  Instead, the evolution is caused by the stationary simultaneous interaction of two heat baths (having distinct spectra and temperatures) with a single two-level system that is in turn coupled to the quantum piston.
 The fully quantized treatment we put forward allows us to investigate work extraction and refrigeration by the tools of quantum-optical amplifier and dissipation theory, particularly, by the analysis of amplified or dissipated phase-plane quasiprobability distributions.
Our main insight is that  quantum states may be thermodynamic resources  and can provide a powerful handle, or control, on the  efficiency of the heat machine. In particular, a piston initialized in a coherent state can cause the engine to produce work at an efficiency above the Carnot bound in the linear amplification regime. In the refrigeration regime, the coefficient of performance can transgress the Carnot bound if the piston is initialized in a Fock state. 
The piston may be realized by a  vibrational mode, as in nanomechanical setups, or an electromagnetic field mode, as in cavity-based scenarios.
\end{abstract}
\maketitle

\section{Introduction}

Thermodynamics and quantum optics have been intertwined since the inception of field quantization \cite{planck00,einstein16}. Over the years their interconnection has been  repeatedly revealed, e.g., in the derivation of the maser efficiency from thermodynamics \cite{ScovilPRL59,GeusicPR67,GeusicJAP59} and its extension to the micromaser \cite{ScullyBOOK97}, in suggestions to boost Carnot-cycle efficiency through bath preparation in nonthermal (population-inverted \cite{Landsbergjpa77,LandsbergJAP80} or ``squeezed'' \cite{rossnagelPRL14}) states, and in the proposed efficiency enhancement of cavity-based amplifiers \cite{ScullySCI03,BoukobzaPRA13} and solar cells \cite{ScullyNAS11} through quantum coherence (interference) effects.

Here we wish to further expand the fruitful rapport between these two disciplines by exploring the thermodynamic capacity for work extraction and cooling of quantum states. To achieve this we study the dependence of the performance bounds of heat machines (engines and refrigerators) on the \textit{initial quantum state} and subsequent evolution of their drive, alias piston, modeled by a quantized harmonic oscillator.The piston may be a mechanical vibrational mode as in optomechanical setups \cite{aspelemeyerarxiv13}, or an electromagnetic field mode, as in masers or lasers \cite{ScullyBOOK97}. Our goal is to provide a fully quantized treatment of \textit{self-contained (autonomous)} heat machines, as opposed to their prevailing semiclassical description:  ``working-fluid'' system, intermittently coupled to heat baths, that is \textit{ parametrically} driven by a classical time-dependent piston or field \cite{AlickiJPA79,GemmerBOOK10,Kosloffentr13,PalaoPRE01,LindenPRL10,LevyPRL12,GevaJCP96,BrunnerPRE12,LevyPRE12,CorreaPRE13,VenturellyPRL13,ScovilPRL59,GeusicPR67,GeusicJAP59,gordonbook00,VelascoPRL97,QuanPRE07,AllahverdyanPRE10,JahnkeEPL10,BirjukovEPJB08,GieselerPRL12,quanpre12,espositoarxiv10,espositopre12,delarxiv13}. The fully quantized treatment we put forward allows us to investigate work extraction and refrigeration by the tools of quantum-optical amplifier and dissipation theory \cite{carmichaelBOOK99}, particularly, by the analysis of amplified or dissipated phase-plane quasiprobability distributions  \cite{Schleichbook01}.

Explicit results are obtained here for a minimal design: it  consists of a ``working-fluid'' realized by a single two-level system (TLS) that  is permanently coupled to two thermal baths with \textit{distinct spectra} and temperatures and is driven by a quantum mechanical harmonic oscillator acting as a piston. We stress that both the TLS and the piston-oscillator are \textit{essential} if the design is to be self-contained, i.e., autonomous: the TLS allows heat flow to or from the baths, i.e., it  mediates between the two baths, and the piston  extracts the work in an engine or provides the energy input in a refrigerator. Namely, the ``working-fluid''  system cannot \textit{directly} extract work or refrigeration from the bath: a piston must be coupled to the system to this end.

%
%
%Since coherence \cite{ScullySCI03,ScullyNAS11} \textit{does not exist} in a  thermalized TLS  that is \textit{weakly} driven by a quantum-mechanical piston, nor do we use a nonthermal  \cite{Landsbergjpa77,LandsbergJAP80,dunkelnatph13} or nonclassical \cite{rossnagelPRL14} bath, we ask:   how is the heat-machine performance affected by the initial preparation of the  piston  in an appropriate quantum state? Our main insights are that the choice of an initial quantum state can provide a powerful handle, or control, on the subsequent performance of the heat machine over finite but long times, i.e., over many cycles  past the equilibration time of the working fluid (TLS).

In order to analyze work extraction or refrigeration in such a \textit{self-contained} quantized setup, we  have to forego   the standard division of energy-exchange between heat, Q, and work, W,  that is known to apply under \textit{classical (parametric)} driving  of the reduced state of the working-fluid system, $\rho_S(t)$, via  a cyclic  Hamiltonian  $H_S(t)$.  This division is expressed by \cite{AlickiJPA79}

\begin{equation}
Q=\oint tr\{\dot{\rho}_S H_Sdt\},
W=-\oint tr\{\rho_S\dot{H}_Sdt\}.
\label{eq:qw}
\end{equation}

Work is maximized when the evolution $\rho_S(t)$ is purely unitary (App. A)

   These standard formulae 
\textit{do not apply} in the present scenario, since $H_S$ is  now time-independent, thus necessitating  an alternative analysis. Our analysis is based on the notion of \textit{non-passivity} of $\rho_S$, which defines its capacity to deliver work, i.e., the maximal amount of work it can yield \cite{LenJSP78,puszcmp1978} (Sec. IIB) By contrast, refrigeration does not involve non-passivity (Sec. IIC).  In Sec. III the evolving bounds on work and refrigeration efficiencies are analyzed for the minimal design discussed above, revealing their crucial dependence on the initial quantum state of the piston. Although Eq. \eqref{eq:qw} does not hold for the fully quantized scenarios considered below, we discuss the correspondence of these two approaches, i.e., the retrieval of the semiclassical Eq. \eqref{eq:qw} and the corresponding results for masers \cite{ScullyBOOK97} (Sec. IV). Possible implementations and their characteristics are discussed in Sec. V.

\section{Work and cooling  in quantized heat machines at steady-state of the system}\label{bound}

\subsection{Basic assumptions and principles}

We assume that the hot and cold  baths (H and C) are only coupled to the ``working-fluid'' system denoted by S, whereas S is coupled to the piston (P). The total time-independent Hamiltonian is then

\begin{equation}
H_{tot}=H_S+H_P+H_{SP}+\sum_j (H^j_{SB}+H^j_{B}),
\label{eq:htot}
\end{equation}
 where $j=H,C$ is the bath index.

The \textit{direct} interaction of S with the baths, forces S to be in a \textit{periodic steady state (limit cycle)} \cite{Kosloffentr13}. Although P is isolated from the baths, they change its energy and entropy \textit{indirectly} via S.  As opposed  to S, P  \textit{cannot be assumed to  maintain constant energy or purity}, because this would be incompatible with its role as a piston  that draws work or cools down the setup. Namely,  the state of the piston must inevitably keep changing and cannot be fully cyclic.  Such a change  of P is not explicitly accounted for  in prevailing  heat-machine models \cite{GemmerBOOK10,Kosloffentr13,PalaoPRE01,LindenPRL10,LevyPRL12,GevaJCP96,BrunnerPRE12,LevyPRE12,CorreaPRE13,VenturellyPRL13,ScovilPRL59,GeusicPR67,GeusicJAP59,gordonbook00,VelascoPRL97,QuanPRE07,AllahverdyanPRE10,JahnkeEPL10,BirjukovEPJB08,GieselerPRL12,quanpre12,espositoarxiv10,espositopre12,ScullySCI03,ScullyNAS11,BlickleNATP11,BoukobzaPRA13}  which assume parametric driving by a \textit{constant classical field}, as in Eq. \eqref{eq:qw}. Here, by contrast, the quantization of P and its initial-state preparation necessitate the consideration of this evolution, as in the theory of  quantum amplifiers (lasers and masers) or dissipators \cite{ScullyBOOK97,carmichaelBOOK99}.

To evaluate  work extraction  or refrigeration  by P, we may invoke energy conservation (the first law of thermodynamics)

\begin{equation}
\mathscr{J}_{H}+\mathscr{J}_{C}-\frac{d\langle H_{P}\rangle}{dt}=0,\label{eq:enecons}
\end{equation}

\noindent where $\mathscr{J}_{H(C)}$ are the heat-flow rates (currents) from H or C to S, respectively, whereas $\frac{d\langle H_{P}\rangle}{dt}$ is  the mean piston-energy change-rate. As discussed below, the common assumption that $\frac{d\langle H_{P}\rangle}{dt}$ represents the rate of  work \cite{GemmerBOOK10,Kosloffentr13,PalaoPRE01,LindenPRL10,LevyPRL12,GevaJCP96,BrunnerPRE12,LevyPRE12,CorreaPRE13,VenturellyPRL13,ScovilPRL59,GeusicPR67,GeusicJAP59,gordonbook00,VelascoPRL97,QuanPRE07,AllahverdyanPRE10,JahnkeEPL10,BirjukovEPJB08,GieselerPRL12,quanpre12,espositoarxiv10,espositopre12,ScullySCI03,ScullyNAS11,BlickleNATP11,BoukobzaPRA13} only holds for classical P, whereas for a quantized P it   is a \textit{combination
of heat flow and work-producing power}. Their ratio strongly depends on the piston state.

Under weak system-bath coupling, $\langle H_P(t) \rangle$ undergoes quasi-cyclic, slowly-drifting evolution (Sec. III) which is the nonadiabatic counterpart of Carnot cycles. The steady-state of S and the slow-changing cycles of P correspond to Markovian evolution of S+P \cite{Gelbwaserumach,LindbladBOOK83,SzczygielskyPRE13}.

  The  \textit{bound}  for the total entropy-production rate of S+P is provided by the Clausius   version of the second law  in the form of Spohn's inequality that holds  under Markovian evolution \cite{SpohnJMP78}. 
	Assuming a  small ratio of the system-piston coupling strength $g$ to the piston oscillation-energy (frequency) $\nu$, the system and the piston are nearly in a product state,  their   production of entropy is even closer to being additive (see App.): 
	
	\begin{equation}
	\rho_{S+P}=\rho_S\otimes\rho_P+O(\frac{g}{\nu})^{2}; \quad
	\dot{\mathscr{S}}_{S+P}=\dot{\mathscr{S}}_S+\dot{\mathscr{S}}_P+O(\frac{g}{\nu})^{4}.
	\label{eq:rhosp}
	\end{equation}

\noindent	Then, considering that after cross-graining,  $\dot{\mathscr{S}}_S=0$ at periodic steady-state and the only entropy production is that of the piston, $\dot{\mathscr{S}}_P$, the second law expressed by the Spohn inequality reads 

\begin{equation}
\dot{\mathscr{S}}_P \geq\frac{\mathscr{J}_{H}}{T_{H}}+\frac{\mathscr{J}_{C}}{T_{C}}.
\label{eq:2law}
\end{equation}

\noindent In what follows, this inequality will be used to infer efficiency bounds that allow for entropy and work production by P.

\subsection{Work efficiency bound with quantized piston}
What is the proper definition of work   when the standard formula \cite{AlickiJPA79} (see Introduction) \textit{does not apply},  since  P   and  S interact via a \textit{time-independent} Hamiltonian, as in Eq. \eqref{eq:htot}? To this end, we invoke in what follows a seldom-used but  rigorous definition of work capacity that is based on the notion of \textit{passivity} \cite{LenJSP78,puszcmp1978} in fully quantized setups.

For a given $\rho_P$, \textit{the work capacity is the maximum  extractable work} (Fig. 1c) expressed by

\begin{equation}
W_{Max} (\rho_P)= \langle H_P(\rho_P) \rangle - \langle (H_P(\tilde{\rho}_P)\rangle
\label{eq:maxw}
\end{equation}

\noindent where  $\tilde{\rho}_P$ is a   state  that minimizes the mean energy of P, without changing its entropy,  and thus \textit{maximizes}  the  work extractable from $\rho_P$. In fact, $\tilde{\rho}_P$ is a  \textit{passive state} \cite{LenJSP78,puszcmp1978}, defined as a state for which  $W_{Max}(\tilde{\rho}_P)=0$ , i.e., a state in which work cannot be extracted. A notable example of a passive state  (out of an infinite variety) is a Gibbs (thermal) state, as discussed below. An equivalent definition of the passivity of a state utilized in Sec. \ref{sec:min} is  a distribution that falloff as the energy increases.

 \cite{LenJSP78,puszcmp1978}: any deviation of the distribution from such monotonicity renders it nonpassive. Nonpassivity is an unambiguous, quantitative measure of the thermodynamic behavior of a quantum state. It will be shown to differ from known characteristics of quantum states, such as their purity or Wigner-function negativity \cite{ScullyBOOK97,Schleichbook01}.

As the initial state of the piston   (P), $\rho_P(0)$, evolves (via a Markovian  master equation \cite{lindbladcmp75})   to a  state $\rho_P(t)$, the  maximum extractable work changes, according to Eq. \eqref{eq:maxw},  by the amount 

\begin{equation}
\Delta W_{Max}(t)=W_{Max} (\rho_P(t))-W_{Max}(\rho_P(0)).
\label{eq:deltaw}
\end{equation}

\noindent To have \textit{positive work production} or  equivalently an increase in work capacity, i.e., $\Delta W_{Max}(t) >0$, it is necessary (but not sufficient) to launch P in a \textit{nonpassive}  state,  $\rho_P(0)$, because a passive state cannot become nonpassive under   Markovian (dissipative or amplifying)  evolution, as shown in Sec. III.

The  \textit{upper bound} for $W_{Max} (\rho_P)$ (to be used in $\Delta W_{Max}$) is obtained  by taking \textit{the lower bound} of  the second term in Eq. \eqref{eq:maxw}, i.e., setting 

\begin{equation}
\langle H_P (\tilde{\rho}_P)\rangle  =\langle H_P \rangle _{Gibbs},
\label{eq:hgibbs}
\end{equation}

 \noindent since  the  Gibbs state 
 is the minimal-energy state with the same entropy as $\rho_P$ \cite{CallenBOOK85,GemmerBOOK10}.
 We stress that
 any $\rho_P$ can be associated  with a  \textit{fictitious} Gibbs state,

\begin{equation}
\tilde{\rho}_P(t)=Z^{-1}e^{-\frac{H_P}{T_P}}
\label{eq:rpt}
\end{equation}
  that has the same entropy, so that we may assign it an \textit{effective} temperature $T_P(t)$. If $\rho_P(t)$ happens to be  a \textit{thermal} state, then $T_P$ is its \textit{real} temperature, otherwise it is merely a parameter that characterizes its evolution.
	
Upon taking the time derivative of this upper bound of Eq. \eqref{eq:maxw}, using Eqs. \eqref{eq:hgibbs}, \eqref{eq:rpt}, we find that the  \textit{extractable power}  is maximized by \cite{gelbiepl13}

\begin{gather}
\mathcal{P}^{Max}=
\frac{d\langle H_{P}\rangle}{dt} -T_P  \dot{\mathscr{S}}_P(t); \quad T_P \dot{\mathscr{S}}_P=\frac{d\langle H_P\rangle_{Gibbs}}{dt},
\label{eq:pnonpas}
\end{gather}

\noindent in terms of the evolving temperature and  the entropy-production rate $\dot{\mathscr{S}}_P$  of the effective Gibbs state.

 Equation \eqref{eq:pnonpas}   yields the quantum heat-engine  (QHE) efficiency bound  in the work-production regime
 
\begin{gather}
\eta^{Max}= 
\frac{\mathcal{P}^{Max}}{\mathscr{J}_H}=
\frac{\frac{d\langle H_{P}\rangle}{dt}-T_P \dot{\mathscr{S}}_P}{\mathscr{J}_H}>0.
\label{eq:pasbound}
\end{gather}

\noindent The  term $-T_P \mathscr{\dot{S}}_P$ on the r.h.s of \eqref{eq:pasbound}, reflecting the heating and entropy change of P,  is neglected by  the prevailing semiclassical treatments that treat P as a classical parametric drive of S  
\cite{GemmerBOOK10,Kosloffentr13,PalaoPRE01,LindenPRL10,LevyPRL12,GevaJCP96,BrunnerPRE12,LevyPRE12,CorreaPRE13,VenturellyPRL13,ScovilPRL59,GeusicPR67,GeusicJAP59,gordonbook00,VelascoPRL97,QuanPRE07,AllahverdyanPRE10,JahnkeEPL10,BirjukovEPJB08,GieselerPRL12,AllahverdyanPRE05,Landsbergjpa77,LandsbergJAP80,dunkelnatph13,rossnagelPRL14,Gelbwaserumach},  but
$ \dot{\mathscr{S}}_P$ cannot be ignored for \textit{a quantum piston},  as shown below. Despite its being ``fictitious'' or effective, the product $T_P \dot{\mathscr{S}}_P(t)$ is a faithful measure of the piston heating rate, because it expresses the rate of its passivity increase (or nonpassivity loss), as illustrated in Sec. \ref{sec:min}.

The compliance of \eqref{eq:pasbound} with the standard Carnot bound is only  ensured if $T_C \leq T_P$. Yet for  $T_P < T_C$,  the Spohn inequality  \eqref{eq:2law} 
 implies   that  Eq.  \eqref{eq:pasbound} satisfies

\begin{gather}
\eta^{Max} (T_P < T_C)\leq 1-\frac{T_P}{T_{H}}.
\label{eq:tpsmall} 
\end{gather}
 
\noindent  The efficiency in   Eq. \eqref{eq:tpsmall} \textit{surpasses} the standard two-bath Carnot bound, $1-\frac{T_C}{T_H}$, when $T_P<T_C$. Nonetheless, Eq. \eqref{eq:tpsmall} adheres to  Spohn's inequality \cite{SpohnJMP78} and therefore to \textit{the second law}.

  Hence, a   quantized treatment of a heat engine yields an efficiency that is \textit{necessarily} determined by  the \textit{effective} temperature $T_P$ of the piston and not only by the real bath temperatures $T_H$ and $T_C$ (provided they are non-negative, as opposed to Refs. \cite{Landsbergjpa77,LandsbergJAP80,LevyPRE12}).
	
	This outcome of the quantized treatment of the piston		
 has a classical analog: a  heat  engine operating between \textit{three} different temperatures has the maximum efficiency, $
\eta=1-\frac{T_{low}}{T_{high}}
$,
 where $T_{low}$ is the lowest and $T_{high}$ is the highest of the three temperatures.  Yet, whereas classically, three real baths are  required, we see that the outcome of our analysis is that a   \textit{single-mode }quantized piston \textit{inevitably} plays the role of the third bath. Its effective temperature $T_P$ \textit{depends on its quantum state},  ranging in the course of time between $T_{low}$ and $T_{high}$.

		Our focus in what follows is on the dependence of  $\eta^{Max}$  on the initial quantum state of the piston.

\subsection{Refrigeration efficiency bound with quantized piston}

We next wish to infer, in the most general form, the conditions for the cooling  of C, i.e. for the quantum refrigerator (QR) regime, which amounts to $\mathcal{J}_C>0$. Substituting \eqref{eq:enecons} in \eqref{eq:2law}   and dividing $\mathcal{J}_C>0$ by the \textit{input} energy flow  from the piston, $-\frac{d\langle H_{P}\rangle}{dt}$, we obtain the upper bound for the coefficient of performance (COP)   of the QR

\begin{equation}
\mathscr{C}=\frac{\mathscr{J}_{C}}{-\frac{d\langle H_{P}\rangle}{dt}}\leq\frac{1}{\frac{T_{H}}{T_{C}}-1}\left(1-\frac{\dot{\mathscr{S}}_{P}T_{H}}{\frac{d\langle H_{P}\rangle}{dt}} \right).
\label{eq:COP}
\end{equation}

\noindent The  first factor on the r.h.s of the inequality is the \textit{standard} bound  (the reciprocal of the Carnot bound): 
$
 COP(\dot{\mathscr{S}}_P=0)\leq\frac{1}{\frac{T_{H}}{T_{C}}-1}.
$
The factor in brackets is the object of interst: for nonpassive  states (which  ``store'' work), we can have 

\begin{equation}
\dot{\mathscr{S}}_{P}/\frac{d\langle H_{P}\rangle}{dt}<0,
\label{eq:difsign}
\end{equation}
  since the entropy of a nonpassive state may increase, $\dot{\mathscr{S}}_P>0$, while $\frac{d\langle H_{P}\rangle}{dt}<0$, so that P may  use its energy (in the form of work) for refrigeration. In this case  the \textit{COP in \eqref{eq:COP}} may exceed the Carnot bound. 
To this end, P must simultaneously receive heat and  deliver power. Only certain quantum states of P possess this ability, as shown below; for those states, the   COP  \textit{ transgresses the standard Carnot bound}. Although a classical analog of this resource is conceivable, our interest  is in \textit{its dependence on the quantum state of the piston}.

 The lower bound of \eqref{eq:COP} is  the maximum efficiency of an \textit{absorption}
QR \cite{GemmerBOOK10} : it  corresponds to a piston that is completely thermalized and delivers heat only, $\frac{d\langle H_{P}\rangle}{dt}=T_P \dot{\mathscr{S}}_P$. Then
 the COP of a QR that is completely driven by heat, becomes
\begin{equation}
\mathscr{C} \geq\frac{1}{\frac{T_{H}}{T_{C}}-1}\left(1-\frac{T_{H}}{T_{P}}\right).
\label{eq:coplowbound}
\end{equation}

The refrigerator COP bounds \eqref{eq:COP}  and \eqref{eq:coplowbound},  and the work-production QHE efficiency bound  \eqref{eq:tpsmall} require very different  conditions. In what follows we inquire: Are they realizable at all, and, if so, can they be attained in the same machine? How do these bounds depend on the initial quantum state of the piston and its evolution?

\section{ Minimal Model Analysis} \label{sec:min}
  To answer the questions raised above, the  general  analysis  presented  in Sec. \ref{bound} will now be  applied to the simplest  (minimal)   model   conforming to  the Hamiltonian in Eq. \eqref{eq:htot}. We shall  assume that
the working-fluid S is a two-level system (TLS) and the  system-bath  (S-B) coupling, $H_{SB}$,  has the  spin-boson form. The TLS  coupling  to the harmonic-oscillator piston (P) complements the Hamiltonian.
The most general   Hamiltonian  of this kind that can yield work can be written as

\setcounter{equation}{0}
\renewcommand{\theequation}{16\alph{equation}}
\begin{gather}
H=H_{S+P}+\sigma_X\otimes(B_{C}+B_{H})+\sum_{j=H,C}H_{B_{j}}; \label{eq:hoa}\\
H_{S+P}=\frac{1}{2}\omega_{0}\sigma_{Z}+\nu a^{\dagger}a+H_{SP}. \label{eq:hob}
\end{gather}

\noindent Here   $B_{H(C)}$ are the multimode bath operators. The $\sigma_X$-coupling in $H_{SB}$ is the only interaction capable of extracting work from the bath via S, as opposed to $\sigma_Z(B_C+B_H)$ coupling that only causes dephasing,  does not contribute to work, because it commutes with the energy of  S and therefore cannot pump energy (from the bath via S) into P.

  The S-P interaction Hamiltonian, $H_{SP}$,   may have one of the following forms:

\noindent i) If the harmonic-oscillator P   is  \textit{off-resonantly (dispersively)} coupled to the TLS,  then

\setcounter{equation}{0}
\renewcommand{\theequation}{17\alph{equation}}
\begin{equation}
H_{SP}=g\sigma_{Z}\otimes(a+a^{\dagger}),
\label{eq:sigmaz}
\end{equation}

\noindent $g$ being  the coupling strength and $a$, $a^\dagger$,  respectively,  the P-mode annihilation and creation operators   \cite{blaisPRA04,xiangRMP13}.  

 \noindent ii)  Alternatively, the spin-boson interaction Hamiltonian may be considered, 

\begin{equation}
H_{SP}=g(\sigma_{+}a+\sigma_{-}a^{\dagger}),
\label{eq:sigmax}
\end{equation}

\renewcommand{\theequation}{\arabic{equation}}
\setcounter{equation}{17}

\noindent where S and P are \textit{ resonantly coupled} and obey the rotating-wave approximation \cite{ScullyBOOK97}.

\subsection{Diagonalization}

The analysis of the model \eqref{eq:sigmaz} or \eqref{eq:sigmax} is simplified by using a new set of canonical
operators $b,b^{\dagger}$ and Pauli matrices $\widetilde{\sigma}^{k}$ $(k=X,Y,Z)$
obtained from $a,a^{\dagger},\sigma_{k}$ by a unitary \emph{dressing transformation} 
that diagonalizes the Hamiltonian.

The transformation
\begin{equation}
a\mapsto b=U^{\dagger}aU,\ \sigma_{k}\mapsto\widetilde{\sigma}_{k}=U^{\dagger}\sigma_{k}U\ ,\ U=e^{\frac{g}{2\nu}(a^{+}-a)\sigma_{Z}}.\label{dress}
\end{equation}

\noindent diagonalizes the Hamiltonian  \eqref{eq:hob} to the form

\begin{equation}
H=H_S+H_P; \quad
H_S=\frac{1}{2}\omega_{0}\sigma_{Z}; \quad H_P=\nu b^{\dagger}b-(\frac{g}{2})^{2}\frac{1}{\nu}.\label{ham_TLSQB_1}
\end{equation}

 \noindent The Pauli matrix $\sigma_{X}$ which appears in the system-bath Hamiltonian
$(H_{SB})$ in \eqref{eq:hoa} is given in terms of  the new dynamical variables as 

\setcounter{equation}{0}
\renewcommand{\theequation}{20\alph{equation}}
\begin{equation}
\sigma_{X}=\widetilde{\sigma}_{+}e^{\frac{g}{\nu}(b^{\dagger}-b)}+e^{-\frac{g}{\nu}(b^{\dagger}-b)}\widetilde{\sigma}_{-}.
\label{eq:sigmaxint}
\end{equation}

\noindent The Heisenberg-picture Fourier decomposition of $\sigma_{X}$ to lowest order in $g/\nu$,
can be obtained in the form

\begin{gather}
\sigma_{+}(t)=e^{i{H}t}{\sigma_{+}}e^{-i{H}t}=e^{i\omega_{0}t}\widetilde{\sigma}_{+}e^{\frac{g}{\nu}(b^{\dagger}e^{i\nu t}-be^{-i\nu t})}\approx \notag\\
\widetilde{\sigma}_{+}e^{i\omega_{0}t}+\frac{g}{\nu}\bigl(S_{1}^{\dagger}e^{i(\omega_{0}+\nu)t}-S_{-1}^{\dagger}e^{i(\omega_{0}-\nu)t}\bigr); \notag\\ 
 S_{1}^{\dagger}=\widetilde{\sigma}_{+}b^{\dagger}\ ,S_{-1}^{\dagger}=\widetilde{\sigma}_{+}b. 
\label{eq:fourier-1}
\end{gather}

\renewcommand{\theequation}{\arabic{equation}}
\setcounter{equation}{20}

The  approximation made above in \eqref{eq:fourier-1} is valid
for the low-excitation regime of the piston which will be shown to correspond to its linear amplification  or dissipation

\begin{equation}
\frac{g}{\nu}\sqrt{\langle b^{\dagger}b\rangle}=(g/\nu) \langle H_P\rangle^{1/2}<<1.\label{regime-1}
\end{equation}

At the  steady-state for S, we are interested in the evolution  of $\langle H_P \rangle$. Since $\sigma_X$, the operator that couples the system to the bath is in Eq. \eqref{eq:hoa}, is seen from \eqref{eq:sigmaxint} and \eqref{eq:fourier-1} to be mixed with $\widetilde{\sigma}_{+}b^{\dagger}$ and $\widetilde{\sigma}_{+}b$, it is clear that $H_P$ in Eq. \eqref{ham_TLSQB_1} will be affected by the system-bath coupling. Namely, the energy exchanged between the system and the baths will be compensated by the energy exchanged between the system and the piston.  In what follow the rates of this energy exchange are analyzed. 

\subsection{Markovian evolution: the Fokker-Planck equation}

We   describe the bath-induced dynamics by the Lindblad generator, which adheres to the second law \cite{LindbladBOOK83}.
 This generator involves the bath response at the Hamiltonian eigenvalues\cite{SzczygielskyPRE13}:   the TLS (S) resonant  frequency, $\omega_0$,  and combination frequencies $\omega_{\pm}$: 

(i) for \eqref{eq:sigmaz} (dispersive $H_{SP}$) 

\begin{equation}
\omega_{\pm}=\omega_0 \pm \nu,  
\label{eq:omegapm}
\end{equation}

where $\nu$ is  the  piston (P) frequency; 

(ii) for \eqref{eq:sigmax} (spin-boson $H_{SP}$)

\begin{equation}
\omega_{\pm}=\nu \pm \frac{g^2}{4\delta},
\label{eq:osb}
\end{equation}

 where $\delta=\omega_0-\nu$.

 For either $H_{SP}$, the corresponding master equation for the state of S+P is

\begin{equation}
{\frac{d\rho_{S+P}(t)}{dt}}=\sum_{q=0,\pm1}(\mathcal{L}_{q,H} +\mathcal{L}_{q,C})\rho_{S+P}(t).
\label{ME_gen}
\end{equation}
 \noindent Here  $q=0,\pm 1$ labels the harmonics  $\omega_0, \omega_{\pm}$, respectively, 
and the generators  associated with these harmonics in  the two baths, $\mathcal{L}_{q}^{j}$ ($j=H,C)$, have  
 the following  Lindblad  form  (upon denoting the bath-response rates by $G_j(\omega_q)$ and setting $\rho \equiv \rho_{S+P}$)

\begin{gather}
\mathcal{L}_{0,j}\rho=\frac{1}{2}\Bigl\{ G_{j}(\omega_{0})\bigl([\widetilde{\sigma}_{-}\rho,\widetilde{\sigma}_{+}]+
[\widetilde{\sigma}_{-},\rho\widetilde{\sigma}_{+}]\bigr)+ \notag\\
G_{j}(-{\omega}_{0})\bigl([\widetilde{\sigma}_{+}\rho,\widetilde{\sigma}_{-}]+[\widetilde{\sigma}_{+},\rho\widetilde{\sigma}_{-})\Bigr\},\label{generator_loc}
\end{gather}
 
\begin{gather}
\mathcal{L}_{q,j}\rho_{tot}=\frac{g^{2}}{2\nu^{2}}\Bigl\{ G_{j}(\omega_{q})\bigl([S_{q}\rho,S_{q}^{\dagger}]+[S_{q},\rho S_{q}^{\dagger}]\bigr)+ \notag\\
G_{j}(-{\omega}_{q})\bigl([S_{q}^{\dagger}\rho,S_{q}]+[S_{q}^{\dagger},\rho S_{q}]\bigr)\Bigr\}\ ,\ q=\pm1.\label{generator_loc1}
\end{gather}

Thus, the time-independent autonomous Hamiltonian yields a  bath-induced evolution of the S+P state, as discussed below. 

Under this bath-induced dynamics, the reduced density matrix $ \rho_{S+P}(t)$  allows us to  compute the heat currents $\mathscr{J}_{C(H)}$  and the effective temperature $T_P$.
 In particular, the cold heat current (from C to S) is then  given by the expression

\begin{gather}
\mathscr{J}_C=  \sum_{q=0\pm 1}\mathrm{Tr}\bigl((H_{S+P})\mathcal{L}_q^C \rho_{S+P} \bigr).
\label{eq:jcdef}
\end{gather}

To investigate the dependence of work and cooling on the state of P in this model we let S reach  \textit{steady-state}.  The  master equation  (ME) for $\rho_P=Tr_S \rho_{S+P}$ is then isomorphic to  a Fokker-Planck (FP) equation \cite{ScullyBOOK97,carmichaelBOOK99,Schleichbook01} (App.).
Namely, the  Lindblad  ME for the slowly-changing piston can be rewritten as  
\begin{equation}
\dot{\rho}_P = \frac{\Gamma +D}{2}\bigl([b, \rho_P b^{\dagger}] + [b \rho_P, b^{\dagger}] \bigr) + \frac{D}{2}\bigl([b^{\dagger}, \rho_P b] + [b^{\dagger} \rho_P, b] \bigr). 
\label{M2}
\end{equation}

 In its  coherent-state basis,  we can always write $\rho_P=\int{d^2 \alpha \mathbf{P}(\alpha) |\alpha \rangle \langle \alpha|}$, where $\mathbf{P}(\alpha)$  is the corresponding  phase-space (quasiprobability) distribution. The FP equation for any  distribution has then  the form

\begin{gather}
\frac{\partial\mathbf{P}}{\partial t}=\frac{\Gamma}{2}(\frac{\partial}{\partial\alpha}\alpha+\frac{\partial}{\partial\alpha*}\alpha*)\mathbf{P}+D\frac{\partial^{2}\mathbf{P}}{\partial\alpha\partial\alpha*}.
\label{eq:FP}
\end{gather}

\noindent Here  $\Gamma$ and D are the drift and diffusion rates, respectively. They  depend on the sum of the cold- and hot-baths response spectra $G(\omega)=\sum_{j=H,C}G_{j}(\omega)$, sampled at the appropriate combination  (cycle) frequencies $\omega_{\pm}$ for the S-P coupling Hamiltonian \eqref{eq:sigmaz} or \eqref{eq:sigmax}. Explicitly, the drift and diffusion rates satisfy

\begin{gather}
\Gamma= \notag\\
\left(\frac{g}{\nu} \right)^{2}\Big((G(\omega_+)-G(\omega_-))\rho_{11}+\notag 
(G(-\omega_-)-G(-\omega_+))\rho_{00} \Big);  \\
 D=\left(\frac{g}{\nu} \right)^{2}\left(\left(G(\omega_-)\rho_{11}+G(-\omega_+)\rho_{00}\right)\right),
\label{eq:gamad}
\end{gather}

\noindent where $\rho_{11}$ and $\rho_{00}$ are the  populations of the upper and lower  eigenstates of $H_S$ .

For work extraction it is required $\Gamma<0$ (gain) and  one tries to minimize the ratio $D/\Gamma$, so that the piston thermalization induced by diffusion sets in as slowly as possible. These conditions are  \textit{reversed} for refrigeration: $\Gamma>0$ (dissipation) and maximal $D/\Gamma$ are required. 

\subsection{ Work extraction dependence on the piston state}

 The  piston mean-\textit{energy (for either gain or loss)} satisfies 

\begin{equation}
 \langle H_P(t)\rangle = \nu  \frac{D}{\Gamma} (1-e^{-\Gamma t}) +e^{-\Gamma t} \langle H_P(0)\rangle.
\label{eq:energy}
\end{equation}

\begin{figure}
	\centering
		\includegraphics[width=0.5\textwidth]{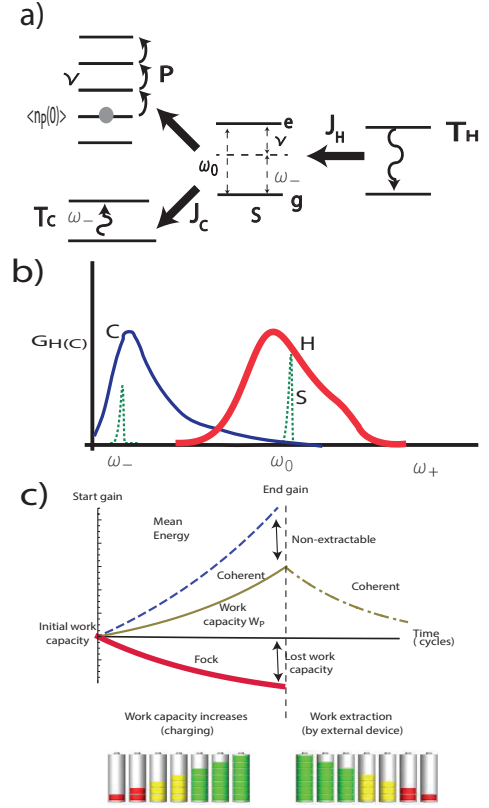}
	\caption{ (Color online) Schematic description of the autonomous quantized heat machine. (a) Scheme of energy and heat exchange between P, C, H and S to elucidate
the heat-pumping gain mechanism of an autonomous quantized heat engine. Reversal of all arrows describes refrigeration via energy investment by P. (b) Spectrally separated  response of the C (thin lines), and H (thick lines) baths  is engineered  using the spectral filtering procedure described in the Discussion. The pointed-line (green) curves) are the $q=-1,0$ ($\omega_-,\omega_0$) harmonics of their response. The $q=1$ ($\omega_+$) harmonic is missing because it does not overlap with either of the (solid) bath spectra: namely, we assume that after the filtering $G_{H}(\omega_{0})\gg G_{C}(\omega_{0})$, and   $G_{C}(\omega_-)\gg G_{C}(\omega_+),G_{H}(\omega_{\pm})$. (c): Schematic drawing of work capacity: ``charging'' the piston in a coherent state by work that is subsequently extracted, upon coupling the piston to an external device.}
	\label{fig:work}
\end{figure}

\noindent This mean energy increases (undergoes gain) 
 for $\Gamma<0$, \textit{ regardless of the passivity (or nonpassivity)  of the initial state.} This \textit{gain represents heat pumping }of P via absorption by  S of a quantum from the H bath at  $\omega_+$ and its emission to the C bath at $\omega_-$ (Fig. \ref{fig:work}a), endowing P with the energy $\omega_+-\omega_-=2\nu$. This process requires  that the two bath-response spectra, $G_H(\omega)$ and $G_C(\omega)$, be \textit{separated} (as in Fig. \ref{fig:work}b), similarly to the semiclassical limit of this model \cite{Gelbwaserumach}.  This condition is however always realizable (see Discussion, Fig. 4c).

\noindent By contrast,   \textit{the work-capacity increase} (expressed by \eqref{eq:deltaw})\textit{ crucially depends upon on the nonpassivity of the initial phase-space 
distribution,} that evolves according to the FP  equation \eqref{eq:FP}.
%This dependence is primarily determined by the  initial state of the piston:

The evolution of any initial distribution is then given by

\begin{equation}
\mathbf{P}(re^{i\theta},t)=\int r_{0}dr_{0}d\theta_{0}\frac{Ke^{-K|(re^{i\theta}-r_{0}e^{-\Gamma t/2}e^{i\theta_{0}})|^{2}}}{\pi}\mathbf{P}(r_{0}e^{i\theta_{0}}),
\label{eq:polar}
\end{equation}

where $\alpha=re^{i\theta}$,  $\alpha_{0}=r_{0}e^{i\theta_{0}}$
and $K(t)=\frac{\Gamma}{D(1-e^{-\Gamma t})}$.
We are interested in its radial derivative,
$
\frac{\partial \mathbf{P}(re^{i\theta},t)}{\partial r}
$
which expresses its passivity or non-passivity, as discussed below for generic cases:

a) If $\Gamma>0$ (\textit{dissipative loss}), then for long times $e^{-\frac{\Gamma t}{2}}\rightarrow0$,
so that

\begin{gather}
\frac{\partial \mathbf{P}(re^{i\theta},t)}{\partial r}=
2r\frac{e^{-\frac{\Gamma r^{2}}{D(1-e^{-\Gamma t})}}}{\pi(\frac{D}{\Gamma}(1-e^{-\Gamma t}))^{2}}<0,\label{eq:gpos}
\end{gather}

\noindent where we used the fact that the distribution is normalized, $\int dr_{0}d\theta_{0}r_{0}\mathbf{P}(r_{0}e^{i\theta_{0}})=1$.
Eq. \eqref{eq:gpos} is \textit{negative} for any $re^{i\theta}$ and \textit{any distribution}.
Hence, for $\Gamma>0$,  any evolving distribution $\mathbf{P}(re^{i\theta},t\rightarrow\infty)$
is passive, and does not allow  work extraction.   

b) Next, assume \textit{an initial passive distribution}, i.e., an \textit{isotropic} distribution satisfying  monotonic decrease with energy: $\frac{\partial \mathbf{P}(r_{0})}{\partial r_{0}}<0$. Then, in the $\Gamma <0$ regime
we find that
$\frac{\partial \mathbf{P}(re^{i\theta},t)}{\partial r}$ is  \textit{negative}, so that $ \mathbf{P}(re^{i\theta},t)$ remains  passive even in the gain regime,  thereby prohibiting \textit{work extraction}.  Hence, \textit{state-passivity is preserved} by the Fokker-Planck phase-plane evolution. 

A notable example of passive-state evolution is that of an initial  thermal state, whose evolution is given by 

\[
\mathbf{P}(\alpha,t|\alpha(0),0)=\frac{\Gamma}{\pi\left(D(1-e^{-\Gamma t})+\sigma\right)}e^{-\frac{\Gamma|\alpha|^{2}}{D(1-e^{-\Gamma t})+\sigma}}
\]

where $\sigma$ is the initial width of the distribution. This state remains thermal (and passive) at any time. 
Although no work is extracted, the mean  energy of \textit{the thermal state
increases for negative $\Gamma$} according to (29). This  example  clearly shows the difference between energy gain and work extraction (Fig, 1c, Fig. 2a). 

c) For an \textit{initially non-passive distribution} in the $\Gamma<0$ regime, we seek the conditions for maximal work extraction. A clue is provided upon introducing the low-temperature approximation to the entropy production rate in Eqs. \eqref{eq:pnonpas},\eqref{eq:pasbound} 

\begin{equation}
\dot{\mathscr{S}}_P(T_p\approx 0) \approx  (\Gamma +2D) (\langle b^\dagger b \rangle -\langle b^\dagger  \rangle \langle b \rangle )+D.
\label{eq:sdotapprox}
\end{equation}

\noindent While the diffusion rate D is a constant, state-independent contribution to entropy-production, the first term is strongly state-dependent as shown in what follows.

\textit{1)  Coherent State:}
An initially  coherent state, $|\alpha(0)\rangle$,  evolves in the gain regime of  the Fokker-Planck equation \eqref{eq:FP}  towards a distribution centered at an exponentially growing $\alpha(t)$ that is progressively broadened by diffusion:

\[
\mathbf{P}(\alpha,t|\alpha(0),0)=\frac{1}{\pi D(1-e^{-\Gamma t})}e^{-\frac{|\alpha-\alpha(0)e^{-\frac{\Gamma}{2}t}e^{-i\nu t}|^{2}}{D(1-e^{-\Gamma t})}}.
\]

In  this case  \textit{there is a unitary operation that transforms this nonpassive  distribution to a Gibbs state, thereby maximizing the work and power extrtaction by the nonpassive state}. This is achieved by displacing  the exponentially growing $\alpha$ towards the origin, by  an amount $\alpha(0)e^{-\frac{\Gamma}{2}t}e^{-i\nu t}$,  thereby attaining the transformed distribution 

\begin{gather*}
\mathbf{\tilde{P}}(\alpha,t|\alpha(0),0) \rightarrow\mathbf{\tilde{P}}(\alpha,t|\alpha(0),0)= \notag \\
\frac{\Gamma}{\pi D(1-e^{-\Gamma t})}e^{-\frac{\Gamma|\alpha|^{2}}{D(1-e^{-\Gamma t})}}.
\label{eq:transp}
\end{gather*}

The  work extraction resulting from such displacement is given by 

\begin{equation}
W_P=\nu|\alpha(0)|^{2}e^{-\Gamma t}
\label{eq:wcoh}
\end{equation}
 
Thus  the coherent-state work capacity  \textit{exponentially increases under  gain  $\Gamma<0$} , as long as our low-excitation  assumption Eq. \eqref{regime-1} holds. Since the entropy-production term is minimized by this state at short times, this  constitutes the optimal case of work extraction. The long-time sustainable work reflects the fact that an initial coherent state retains its nonpassivity and is  never fully thermalized. The geometric condition that underlies work-capacity maximization amounts to keeping the distribution peaked as far away from the origin as possible, so as to maximize its nonpassivity.

\begin{figure}
	\centering
		\includegraphics[width=0.5\textwidth]{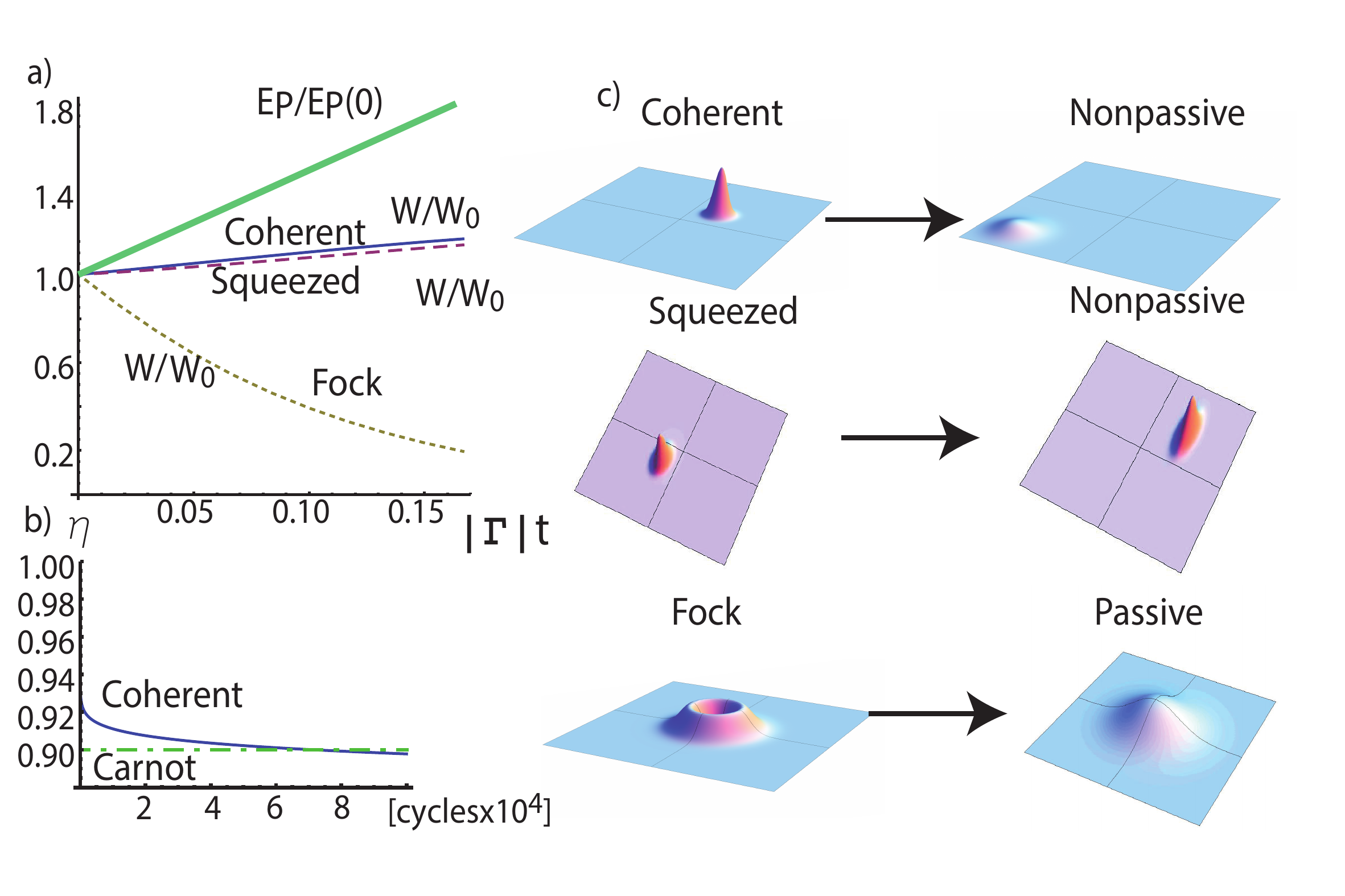}
	\caption{ (Color online) Work capacity evolution. (a) Comparison of work-capacity change (normalized by the initial work capacity) as a function of $|\Gamma|t$ for $E_P(t)/E_P(0)$ (solid thick curve), coherent, quadrature-squeezed  and Fock states of P with the same initial mean energy $E_P(0)=\nu \langle n_P(0) \rangle$. Note their same mean-energy gain. (b) The  efficiency bound of an initial coherent state may surprass  the Carnot limit and remain above it well after the system has reached steady-state (here after $10^4$ cycles). (c) The respective evolution of the phase-plane distribution shows that the high resilience of the initial coherent state against thermalization, as opposed to the lower resilience of quadrature-squeezed states and the fragility of a Fock state, is the key to their different work capacity evolution. The spectral separation conditions and the response harmonics are as in Fig. \ref{fig:work}b to ensure gain in Eq. \eqref{eq:gamad}.}
	\label{fig:effi}
\end{figure}

  For $|\alpha(0)_P|\sim 1$ the efficiency bound $\eta_{Max}$ corresponding to \eqref{eq:wcoh} (Fig. \ref{fig:effi}b) may \textit{exceed} the standard Carnot bound, owing to  the  slow rising entropy and effective temperature $T_P$.
Progressively, the efficiency drops according to  \eqref{eq:tpsmall}, since the effective temperature of $\rho_P$ rises due to diffusion, as

\begin{equation}
1/T_P=\frac{Log(\frac{1+D t}{Dt})}{\nu}.
\label{eq:tpforef}
\end{equation}

Nevertheless, an initial low amplitude coherent state  allows to extract work over many cycles with an efficiency \textit{above the standard two-bath Carnot bound} $1-\frac{T_C}{T_H}$ (Fig. \ref{fig:effi}b).
 The quasiclassical  limit  $ |\alpha(0)_P| \gg 1$ retrieves the  standard Carnot bound, (see Sec. IV).

2) \textit{An initial quadrature-squeezed state} produces higher entropy and therefore  yields less work than a coherent-state with the same mean energy. The same is true for the initial Schrodinger-cat state $\frac{|\alpha_0\rangle+e^{i\theta}|-\alpha_0\rangle}{\sqrt{2}}$ (Fig. \ref{fig:effi}a). This indicates that the\textit{ maximal resilience of a coherent} state against thermalization (passivity) is the key to its higher work efficiency.

\textit{3) An initial  Fock state ($\rho_P(0)=|n_P><n_P|$)} evolves for the relevant times, $D t \ll 1$,  and upon   rescaling  to  $|\alpha'|^2=\frac{\Gamma|\alpha|^{2}}{D(1-e^{-\Gamma t})}$, to the distribution

\begin{gather*}
\mathbf{P}(\alpha',t|\alpha(0),0) \approx e^{-|\alpha'|^2}(1+\frac{l}{D t-1}|\alpha'|^2).
\end{gather*}

For  $|\alpha'|^2 \leq 1$ this distribution is passive,  decreasing as a function of the energy, $|\alpha'|^2$. 
While an initial Fock state of P has some work capacity, once connected to the engine  its work content  does not increase, but rather decreases until it reaches a thermal state. Namely, a Fock state   always evolves to  a passive state.
 In contrast  to the robust coherent state,  the highly fragile  \textit{Fock state} $| n(0)_P \rangle$  quickly thermalizes: it
\textit{ cannot extract  work} from  the  engine and its initial work capacity, $\nu n_P(0)$, is diminished by the engine action (Fig. \ref{fig:effi}a). Therefore $\Delta W_{Max}(t)<0:$ for any time duration,   work production by a Fock state is always \textit{negative} (Fig. 1c).

We have thus reached a central conclusion of this paper: as opposed to energy gain,  the extractable work strongly depends
on the initial  phase-plane distribution  of the piston (Fig. \ref{fig:effi}c):
If initially $\mathbf{P}(\alpha)$ is centered at the origin,  as in a Fock state, it will   rapidly  become passive  for any $\Gamma$ and thereby  terminate work extraction.  A coherent state, by contrast, whose   distribution under gain  $\Gamma<0$ is centered at a growing distance from the origin $|\alpha(t)_P|=|\alpha (0)_P|e^{-\frac{\Gamma t}{2}}$,  will \textit{increase} its nonpassivity, profiting from low entropy production for $|\Gamma|\gg D$, and thus sustain highly efficient  work extraction. Other nonpassive distributions, such as squeezed states or Schrodinger-cat states, undergo faster entropy production. This behavior will persit as long as the linear amplification regime holds   Eq.\eqref{regime-1}, i. e., until the onset of saturation for large $\alpha(t)_P$.

\subsection{ Refrigeration dependence on the piston  state}

The cold   heat current \eqref{eq:jcdef}
 in the model of \eqref{eq:sigmaz} assumes the form 

\begin{gather}
\mathscr{J}_C \propto 
\left(e^{-\frac{\omega_-}{T_{C}}}\langle n_P(t)\rangle-e^{-\frac{\omega_{0}}{T_{H}}}\langle n_P(t)+1\rangle\right),\label{eq:coldJ1}
\end{gather}

\noindent provided that  H and C are spectrally separated, as in Fig. \ref{fig:work}b.

\noindent Here $\langle n_P(t)\rangle$ is the mean number  of  P quanta.  The QR condition $\mathscr{C}>0$ (see \eqref{eq:COP}) then  assumes the form 

\begin{gather}
0< \frac{\omega_-}{\nu}< \frac{1}{\frac{T_H}{T_C}-1};\quad \notag\\
\bar{n}_{min}<\langle n_P(t)\rangle; \quad  \bar{n}_{min}=\frac{1}{e^{\frac{\omega_0}{T_H}-\frac{\omega_-}{T_C}}-1}.
\label{eq:popcond1}
\end{gather}

\noindent Hence,   QR action is limited to $\langle n_P \rangle>\bar{n}_{min}$.  

  The  energy stored in the initial state of the piston  can be used for cooling, provided there is a \textit{a positive drift} in \eqref{eq:FP} $\Gamma>0$ (dissipation). The resulting cooling strongly depends on the  initial states of  P (Fig. \ref{fig:copfin}): 

(a)  For an \textit{initial pure state}, the work stored in P is   equal to its  energy, so that   P
is the  work source of the QR. The refrigerator stops
cooling when Eq. \eqref{eq:popcond1} no longer holds, having  provided  an amount $\nu\langle n_P(0)\rangle$
of work. At the same time, P, initially at $T_P=0$, absorbs
heat from the refrigerator, and ends up in a thermal
state with the critical temperature

\begin{equation}
 (T_{P})_{crit}=T_H\frac{\nu }{\omega_0}\frac{1}{1-\frac{\omega_-}{\omega_0}\frac{T_{H}}{T_C}}.
\label{eq:tp}
\end{equation}

\noindent Thus $(T_{P})_{crit}$ strongly depends on $\frac{\nu }{\omega_0}$. It is noteworthy that   $(T_P)_{crit}\rightarrow \infty$ if the upper bound of \eqref{eq:popcond1} holds. 

The analysis of    \eqref{eq:FP} shows that the COP  in \eqref{eq:COP} does not only depend on the \textit{purity}, but also on the \textit{type} of the initial quantum state:

(i) For an initial \textit{coherent state},  

\begin{equation}
\frac{\dot{\mathscr{S}}_{P}}{\frac{d\langle H_{P}\rangle}{dt}}=D/\left(\left(D-\Gamma \langle n_P(0)\rangle\right)T_P\right).
\label{eq:coheent}
\end{equation} 

 \noindent Namely, its COP is determined by  \textit{the diffusion rate}, $D$, \textit{the drift rate},   $\Gamma$,  and the initial population $ \langle n_P(0)\rangle$.
Denoting $d(t)\equiv\frac{D}{\Gamma}(1-e^{-\Gamma t})$, we find for a coherent state

\begin{gather}
\frac{1}{T_P}= \frac{1}{\nu} \log \left(\frac{1+d(t)}{d(t)}\right), \notag\\
\dot{\mathscr{S}}_P= \nu \frac{D e^{-\Gamma t}}{T_P}.
\label{eq:tpspcoh}
\end{gather}

 (ii) For an initial number (Fock) state  $n_P(0)$, the COP may  \textit{surpass that of a coherent state with the same} $\langle n_P(0) \rangle$
\textit{(Fig. \ref{fig:copfin}a}). The reason is that a Fock state \textit{is much more   prone to thermalization} (Fig. \ref{fig:copfin}b). This allows it to both deliver power  and absorb heat (Fig. \ref{fig:copfin}b)  to a higher degree  than a coherent state, as discussed following \eqref{eq:COP}. Contrary to work extraction, cooling requires the maximization of heat absorption ($\dot{\mathscr{S}}_P$).  Among all the pure states, Fock state has the largest heat absorption capacity, as can be seem from Eq. \eqref{eq:sdotapprox}. 

(iii) Other types of nonclassical initial states, such as Schroedinger-cat states  \cite{ScullyBOOK97} yield a COP  (Fig. \ref{fig:copfin}a) that  is \textit{below that of a Fock state}  with the same $\langle n_P(0) \rangle$, but above that of the corresponding coherent state.

\textit{b) In an initial thermal state,} since the work capacity  is zero, using \eqref{eq:2law} for QR action implies that $T_P>T_H$: P should then be hotter than H.
 In the course of its evolution,  P will remain  in a thermal state, but  $T_{P}$ will decrease until it attains $(T_P)_{crit}$ and  stops cooling.  
Explicitly, an initial thermal state yields

\begin{gather}
\frac{1}{T_P}= \frac{1}{\nu} \log \left(\frac{1+d(t)}{d(t)}\right), \notag\\
\dot{\mathscr{S}}_P= \nu\frac{(D-\Gamma \langle n(0) \rangle_P) e^{-\Gamma t}}{T_P}.
\label{eq:tpspth}
\end{gather}

The corresponding maximal $COP$ conforms to that of an absorption refrigerator \eqref{eq:coplowbound}, which is always lower than the  COP of a power (work) driven refrigerator \cite{LevyPRL12,CorreaPRE13}.

\subsection{Spectrally separated baths}

 The coupling of the two-level system (TLS)  \textit{to spectrally separated baths} (as in Fig. \ref{fig:work}b)  is achievable  by  \textit{spectrally filtered, local} heat-pump
and heat-dump.  The filter can be realized by coupling the TLS through a harmonic-oscillator mode of frequency $\omega_j$ ($j\in H,C$) (``the filter'') to the bath. In this way the TLS becomes \textit{effectively} coupled to two   baths  with  response spectra \cite{KofmanJMO94} (Fig 4(c))

\begin{gather}
G_f^j(\omega)= \notag \\
\frac{\gamma_f}{\pi}
\frac{(\pi G^j(\omega))^2}
{(\omega-(\omega_f^j+\Delta_L^j(\omega)))^2+
(\pi G^j(\omega))^2}, \quad (j=H,C)
\label{eq:efeg}
\end{gather}
 
 where $\gamma_f^j$ is the coupling rate of the TLS to the filter mode, $G^j(\omega)$ is the  original coupling spectrum (without the filter), and 
 
 \begin{equation}
 \Delta_L^j(\omega)=P(\int_0^{\infty} d\omega' \frac{G^j(\omega')}{\omega-\omega'})
\label{eq:DElta}
\end{equation}
 P being the principal value, is the respective bath-induced Lamb shift \cite{KofmanJMO94,cohentBOOK97}. Thus, for any given spectrum $G^j(\omega)$ we obtain
 the effective response spectrum \eqref{eq:efeg} which is a  ``skewed Lorentzian'' with controllable width and center. These parameters may thus be chosen to avoid unwarranted overlap between the  coupling  spectra of the two baths. 

\section{Retrieval of the Semiclassical  results}

\subsection{Semiclassical limit of the nonpassive work extraction}

The fully quantum autonomous heat engine whose work extraction  is determined by nonpassivity should be able to reproduce in the semiclassical limit the power extraction of an externally (\textit{parametrically}) modulated heat engine (proposed in \cite{Gelbwaserumach}) that obeys the cyclic work definition [14] (Eq.(1)), and is governed , instead of $H_{S+P}$ in \eqref{eq:hob} and \eqref{eq:sigmaz} by the Hamiltoninan

\begin{equation}
H_S(t)=\frac{1}{2}(\omega_0+\lambda \nu Sin \nu t)\sigma_z
\label{eq:hst}
\end{equation}
 
while the coupling to the baths is still given by \eqref{eq:hoa}

The two descriptions coincide  when the initial state of the quantum piston is a large-amplitude coherent state $|\alpha|\gg1$.  The extracted power by a piston in an initially coherent state piston  is (Eq. (35))  $-\nu |\alpha_0|^2\Gamma$. For an externally modulated engine, (governed by \eqref{eq:hst}), it  is $-\nu \frac{(\nu\lambda)^2}{(2g)^2} \Gamma$ \cite{Gelbwaserumach}. Thus, a parametric modulation amplitude $\lambda= \frac{2g}{\nu}|\alpha_0|$ provides the same power extraction as the nonpassive coherent state $|\alpha_0 \rangle$.

\subsection{Comparison to the maser }
It is instructive to compare our analysis to that of  the maser, which has long been treated as a heat machine \cite{ScovilPRL59,GeusicPR67,GeusicJAP59,ScullyBOOK97}: its thermodynamic efficiency is given by the ratio of the output (signal) and pump frequencies, which is the same as the Carnot bound. In our scenario there is  no population inversion in the system, and instead the gain is provided by the hot bath, but the analogy is complete. We may recover the maser-gain result for quasiclassical coherent-state preparation of the piston, which is the counterpart of the maser-output (signal) mode, upon substituting the hot bath for the maser-pump mode. Yet there is currently no analysis of the maser efficiency dependence on its initial state, let alone any indication that it may lead to above-Carnot efficiency, since such analysis should rely on the use of nonpassivity for work capacity, as detailed above.

For an initial coherent state  
with large $|\alpha_{0}|^2$ (semiclassical limit) and spectrally separated
baths, such that only $G^{C}(\omega_{0})$ and $G^{H}(\nu_{+})$ are
non-zero, we find  for $|\Gamma| t\ll 1$,

\begin{gather}
\mathscr{J}_{H}\approx\frac{g^{2}}{\nu^{2}}|\alpha_{0}|^2\nu_{+}G^{H}(\nu_{+})G^{C}(\omega_{0})(e^{-\frac{\nu_{+}}{T_{H}}}-e^{-\frac{\omega_{0}}{T_{C}}}); \quad \notag \\
\mathscr{P}=-\Gamma e^{-\Gamma t}|\alpha_{0}|^2\nu\approx-\Gamma|\alpha_{0}|^2\nu
\end{gather}

where 

\begin{equation}
\Gamma\simeq-\frac{g^{2}}{\nu^{2}}G^{H}(\nu_{+})G^{C}(\omega_{0})(e^{-\frac{\nu_{+}}{T_{H}}}-e^{-\frac{\omega_{0}}{T_{C}}}).
\end{equation}

The corresponding efficiency bound becomes 
\begin{equation}
\eta\equiv\frac{\mathscr{P}}{\mathscr{J}_{H}}; \quad \eta^{Max}=\frac{\nu}{\nu_{+}},
\label{eq:efidef}
\end{equation}

\noindent which is the same as in the maser model \cite{ScovilPRL59}.

\begin{figure}
	\centering
		\includegraphics[width=0.5\textwidth]{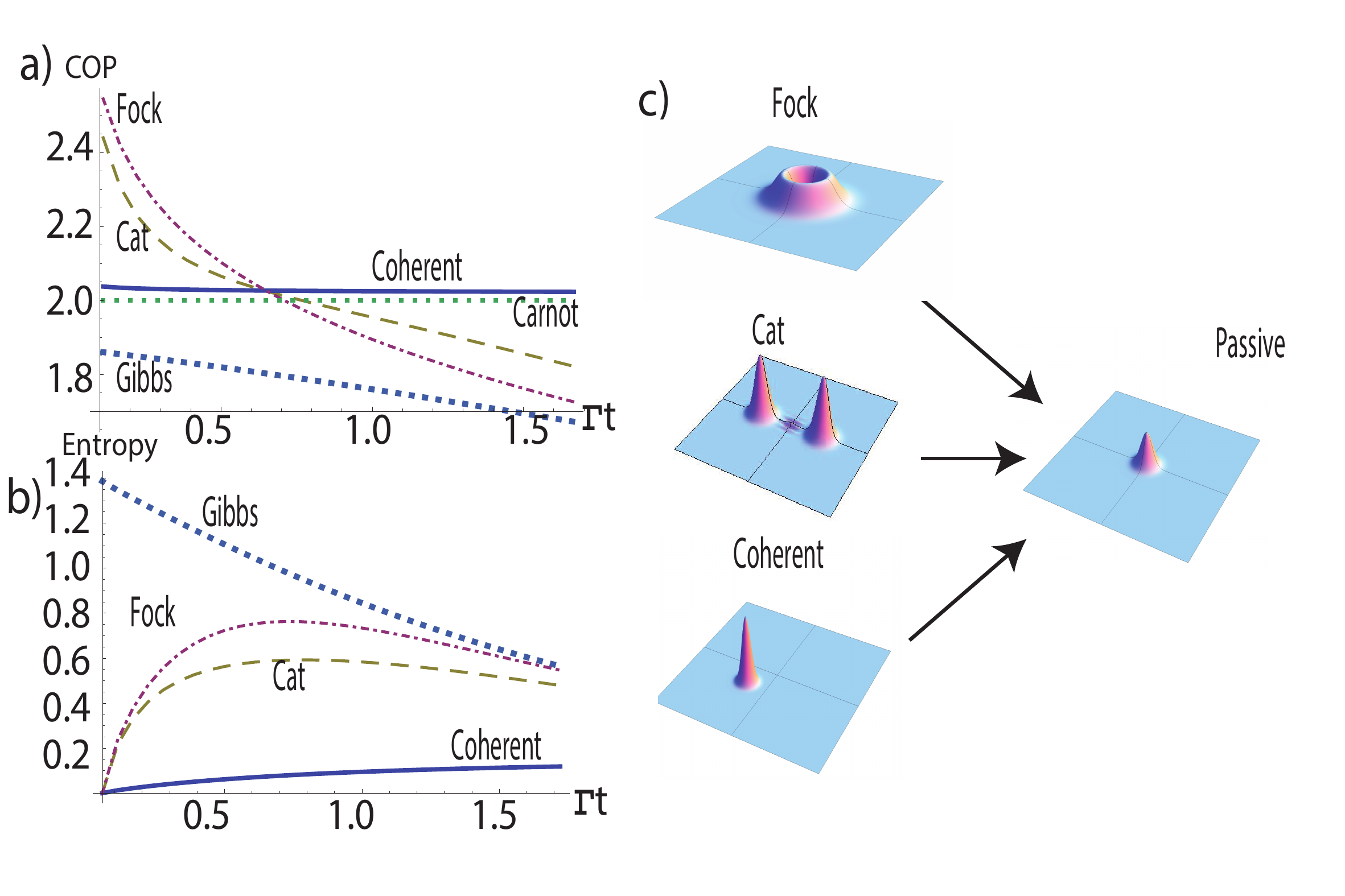}
	\caption{ (Color online) Refrigeration. (a) COP of an autonomous QR as a function of time (in units of $\Gamma^{-1}$) for different initial piston states: coherent (continuous), Fock  (dot-dashed) and thermal  (dashed). The Carnot COP bound (dotted, thin) is  slightly  below the coherent-state COP.  Both thermal-state and Fock-state COP asymptotically coincide to the the absorption bound  at $T_{crit}$ . While at short times the Fock state has the largest COP, the coherent state outperforms it at long times, with a COP  above the  Carnot efficiency.  (b) Entropy (in units of $k_B$) of the same states as a function of time. The initially rapid entropy increase of the Fock state explains its high COP at short times, while for  the coherent state the entropy increase  is slow but steady, resulting in the highest COP at large times. (c) Phase-plane plots of the distribution evolution illustrate that the high resilience of a coherent state against thermalization, as opposed to the low resilience (fragility) of Fock and Schroedinger-cat states, explains the COP evolution. The spectral separation conditions are as in Fig. \ref{fig:work}b, but the piston frequency is chosen to ensure cooling, ($\Gamma>0$). The initial mean energy complies with Eq. \eqref{eq:popcond1}.}
	\label{fig:copfin}
\end{figure}

\section{Discussion}

The present paper has formulated a comprehensive, unified framework for the analysis of   heat-engine and refrigerator thermodynamics under the  hitherto unexplored  quantized-piston drive, both generally and for the simplest (minimal) autonomous model consisting of a two-level system coupled to a quantized-oscillator piston:

1)\textit{Work-extraction capacity by a quantum  piston } is described in terms of  its \textit{deviation  from a passive state}. 
We stress that work extraction  as   defined by the nonpassivity of the piston state, is the only rigorously justifiable measure of work  under \textit{time-independent} Hamiltonian action \cite{LenJSP78,puszcmp1978}. It is shown here to crucially   depend on the initial quantum state,\textit{ in contrast to mean-energy gain. }
 The resulting efficiency bound (Eqs. \eqref{eq:pasbound}, \eqref{eq:tpsmall}) involves  the  effective temperature $T_P$. As long as $T_P <T_C$,  Eq. \eqref{eq:tpsmall}   may surpass the \textit{standard  two-bath }Carnot bound $1-\frac{T_C}{T_H}$, which for conventional masers \cite{ScullyBOOK97,ScovilPRL59} amounts to the  Scovil-Schultz-Dubois output-to-input frequency ratio.  Because it complies with Spohn's inequality \cite{SpohnJMP78},  the present efficiency bound is  consistent with the second law. It shows that  the piston may  serve as a \textit{low-entropy resource} excluded by  the standard (classical-parametric) limit of work extraction.   The cost of such preparation in terms of energy and entropy \textit{does not invalidate} the extra efficiency obtained quantum mechanically in the gain regime. As a consistency check we have shown that the nonpassive work-capacity rate (power) coincides in the semiclassical limit with that of a parametrically-driven engine \cite{Gelbwaserumach} that obeys the standard cyclic-work definition (Sec. IVA).

 One should keep in mind that the Carnot bound of a two-bath heat machine strictly applies only for a zero-entropy piston \cite{CallenBOOK85,GemmerBOOK10,Kosloffentr13,AlickiJPA79,GevaJMO02,ScullySCI03}. By contrast, the  efficiency derived by us is valid  upon allowing for the inevitable but commonly ignored piston entropy growth  and its linear amplification at finite times,  at the steady-state of the system coupled to the baths. This efficiency is both practically and conceptually important for elucidating heat-engine principles in the quantum domain: in this domain,  the initial \textit{``charging'' of the piston } by quantum state-preparation (Fig. 1c) is sought to be maximally efficient. In this respect, our analysis has yielded nontrivial results: (a) in particular, it is remarkable \textit{that as the initial coherent amplitude of the piston decreases, the resulting efficiency increases}, although the \textit{ entropy growth} of the piston might then be expected to reduce (rather than enhance) the efficiency. (b) Work extraction obtained from an initial coherent-state has been found to be superior that of   other states, because of its larger sustainable nonpassivity, conditioned on its low entropy production: equivalently, this reflects the fact that the coherent state is the ``pointer-state'' of the evolution \cite{zurekPRD81}.

2) Our analysis has also  produced several important conclusions concerning  \textit{autonomous quantized refrigerators}, both generally and in the same setup as the heat engine analyzed above: 

(a) While the  semiclassical limit of piston dynamics coincides with that of an externally (classically) driven  QR, once the piston is also quantized, the QR action is dramatically modified:  it proves  \textit{advantageous for the   quantum piston  to increase its temperature and entropy}  during the process, so that its  energy exchange with the system is a combination of heat and work production. The instability (rapid entropy increase) of an initial Fock state compared to an initial  coherent state makes Fock states temporarily superior in terms of cooling. Remarkably, the COP for an initial Fock state exceeds the classical (Carnot) COP, albeit  at short times  (Fig. \ref{fig:copfin}a).

(b) The  quantized piston, once charged, does not require any external
energy source to drive the QR, until it runs out of energy. This could be relevant in situations where 
 external power  is scarce or where miniaturization
of the power source  is an advantage. 

(c) This autonomous QR may act in dual mode:
 The work   stored in an initial  pure state    of the piston may be used to run the machine as a  Carnot QR  (driven by work), as long as the piston is below the critical temperature  \eqref{eq:tp}.
Alternatively, an initial  thermal state of the piston may drive the machine as an absorption QR, provided the piston temperature is above \eqref{eq:tp}.

\begin{figure}
	\centering
		\includegraphics[width=0.5\textwidth]{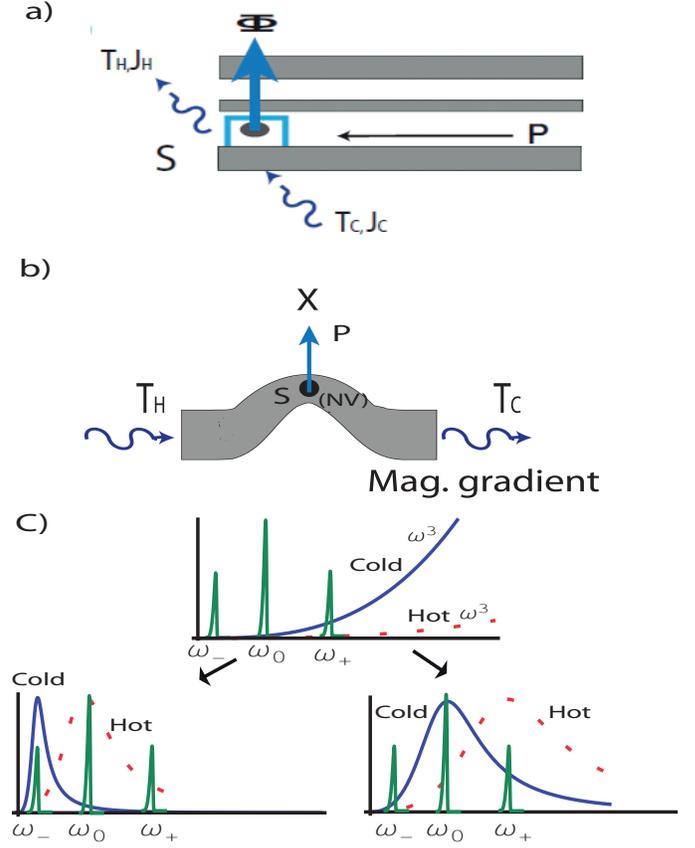}
	\caption{ (Color online) Realizations. (a) Realization of the model in Fig. \ref{fig:work}a, Eq. \eqref{eq:sigmaz}, for a flux qubit in a coplanar resonator. The change in the resonator field (P) affects the magnetic flux threading the qubit. (b) Realization  of the same model for a NV -center defects in diamond mounted on a nanomechanical resonator. The strain-field P affects S that is subject to magnetic-field gradient. (c) Filtering two bath spectra (top) according to  Eqs. \eqref{eq:efeg}, \eqref{eq:DElta}   in order to allow work extraction (bottom right) or refrigeration (bottom, right), by adjusting the filtered bath-spectra overlap with $\omega_0$ and its sidebands in the coupled  S-P system.}
	\label{fig:realizations}
\end{figure}

The  S-P (system-piston) coupling \eqref{eq:sigmaz}  is realizable e.g. for a  superconducting flux   qubit that is off-resonantly   coupled to P, a high-Q  microwave  cavity mode (Fig. \ref{fig:realizations}a) \cite{blaisPRA04,xiangRMP13} .
 Another  design may be based on a NV defect subject to a magnetic-field gradient: the defect is then  coupled to the strain-field (phonon mode) of a nanomechanical resonator (Fig. \ref{fig:realizations} b) \cite{bennettprl13}.  Both setups may attain large gain $|\Gamma| t_{lea} \gg 1$, where $t_{lea}\geq (\frac{\nu}{Q})^{-1}$ is the P-mode leakage time \cite{PetrosyanPRA09,blaisPRA04}.  This allows P to evolve from its initial state to the final state  at a rate $\Gamma$, before it leaks out of the cavity.

To conclude, this  research   reveals  quantum aspects of work and refrigeration, outside the scope of studies which have thus far been  primarily restricted to semiclassical  heat machines i.e., machines driven by an external classical field, whose energy or entropy changes in the course of work extraction on  cooling are imperceptible.  On the applied side, this study is expected to lay the ground for  the design of maximally-efficient  machines powered by work and heat that are stored in  \textit{autonomous}  quantum devices, facilitating  their miniaturization.  On the foundational  side, it  provides deeper understanding of the rapport between thermodynamics and quantum mechanics, as it is concerned with the simplest conceivable,  fully quantized model of a refrigerator or heat engine and shows the existance of thermodynamic resources in quantum states. In particular, it reveals the role of quantum-state nonpassivity and its preservation  under the Fokker-Planck evolution of the piston in the phase-plane for  work-extraction efficiency or, conversely, the loss of nonpassivity for refrigeration efficiency.  

 Effects of quantum  coherence in multilevel media\cite{ScullyNAS11} or entanglement in  multipartite systems \cite{RaoPRL11,kurizkiPRA96,scullysci10,MazetsJPB07,Dillenschneiderepl09,shahmoonPRA13rddi}, as well as effects of engineered non-Markovian environments \cite{ClausenPRL10,KofmanJMO94,ErezNAT08,GordonNJP09,AlvarezPRL10,JahnkeEPL10,GelbwaserPRA13}, may elucidate additional aspects of this fundamental rapport.  The third law \cite{Nernst1912,KolarPRL12,LevyPRL12} should also be revisited in the framework of this model.  Finally, consideration of  quantum fluctuation effects \cite{rossnagelPRL14,seifertRPP12,JarzynskiPRL04,huberprl08} may open a new intriguing avenue of research, wherein initial quantum fluctuations of the piston are related to work or cooling time-dependent fluctuations, beyond their mean values considered in this research.

\textbf{Acknowledgments}
The support of ISF, BSF, AERI, CONACYT is acknowledged.

%\bibliographystyle{ieeetr}
%\bibliography{propbib}

\onecolumngrid
\section*{Appendix}

\setcounter{equation}{0}
\renewcommand{\theequation}{S\arabic{equation}}\

\subsection{Work under adiabatic evolution}

During any unitary process  the entropy remains constant, rendering the process  thermally adiabatic. As shown below  the entire energy change can then be considered  as work.

 Assume that the unitary process is given by  $U=e^{\int_0^{\tau} H_S(t')dt'}$ where $H_S(t')$ is  the Hamiltonian of the system in the Schroedinger picture, which may be controlled by time-dependent external fields.  Then the change of energy during the process is
\begin{equation}
\Delta E_S = \int_0^{\tau} Tr\rho_S(t') \dot{H}_S(t')dt'+\int_0^{\tau} Tr\dot{\rho}_S(t') H_S(t')dt'=\\
\int_0^{\tau} Tr\rho_S(t') \dot{H}_S(t')dt'=W
\label{eq:deltae}
\end{equation}

Upon inserting  the expression for $\dot{\rho}_S(t')=-i [H_S(t'),\rho_S(t')]$ and calculating the trace, the second term on the R.H.S, which is  identified with \textit{heat production} Q, is seen to vanish. Hence, unitary evolution yields \textit{purely work}

\begin{equation}
\Delta E_S = 
\int_0^{\tau} Tr\rho_S(t') \dot{H}_S(t')dt'=W
\label{eq:}
\end{equation}

The purpose of our investigation is to extend this analysis to fully quantized, time-independent Hamiltonians.

\subsection{Entropy production}

We assume that the effect of $H_{SP}$ is a perturbation of strength $\epsilon \ll1$
The correlations
between the system and the piston arise only to second order
in $\epsilon$. Thus the joint state of the system and the
piston can be written as 

\begin{equation}
\rho_{SP}=\rho_{S}\otimes\rho_{P}+\epsilon^{2}\rho_{cor}
\label{eq:totst}
\end{equation}

where $\rho_{cor}$ denotes the correlated part of the joint state. Using  Weyl's inequality, it is possible to estimate the deviation
of the $\rho_{SP}$ eigenvalues from $\rho_{S}\otimes\rho_{P}$, as

\begin{equation}
\lambda_{SP}^{i}=\lambda_{S}^{i}\lambda_{P}^{i}+\epsilon^{2}\lambda_{cor}^{i}
\label{eq:eigen}
\end{equation}

where $\{\lambda_{S}^{i}\},\{\lambda_{P}^{i}\}$ and $\{\lambda_{SP}^{i}\}$ correspond
to the $\rho_{S},\rho_{P}$ and $\rho_{SP}$ eigenvalues respectively.

The Von Neumann entropy of a density matrix $\rho$ is $\mathscr{S}(\rho)=-Tr\rho Ln\rho=-\sum_{k}\lambda^{k}Ln\lambda^{k}$,
where $\lambda^{k}$ are the $\rho$ eigenvalues. Using the fact that $\sum_{k}\dot{\lambda^{k}}=0$
( the trace of a density matrix is always 1), the entropy production is $\dot{\mathscr{S}}(\rho)=-Tr\dot{\rho}Ln\rho=-\sum_{k}\dot{\lambda^{k}}Ln\lambda^{k}$.
At the steady state for the system,  $\dot{\lambda}_{S}=0$, but we still have $\dot{\lambda}_{P},\dot{\lambda}_{cor}\sim O(\epsilon^{2})$. Then 

\begin{equation}
\dot{\mathscr{S}}(\rho_{SP})=\sum_{i}(\lambda_{S}^{i}\dot{\lambda_{P}^{i}}+\epsilon^{2}\dot{\lambda_{cor}^{i}})Log(\lambda_{S}^{i}\lambda_{P}^{i}+\epsilon^{2}\lambda_{cor}^{i})=\sum_{i}\lambda_{S}^{i}\dot{\lambda_{P}^{i}}Log\lambda_{S}^{i}\lambda_{P}^{i}+O(\epsilon^{4})\approx\dot{\mathscr{S}}(\rho_{P})
+O(\epsilon^4)
\label{eq:sdot}
\end{equation}

where the Taylor expansion was used to expand $Log(\lambda_{S}^{i}\lambda_{P}^{i}+\epsilon^{2}\lambda_{cor}^{i})$. Notice that
due to $\dot{\lambda}_{P}\sim O(\epsilon^{2})$, $\dot{\mathscr{S}}(\rho_{P})\sim O(\epsilon^{2})$.

\subsection{From Lindblad evolution to the Fokker-Planck equation}

The  evolution equations obtained from \eqref{generator_loc}, \eqref{generator_loc1}    in the Fock-state basis of P have the from 

\begin{gather}
\dot{\rho}_{nm}^{11}=\sum_{a}-\left(\frac{g^{2}}{2\nu^{2}}\left(G_{j}(\omega_{-})(2+n+m)+G_{j}(\omega_{+})(m+n)\right)+G_{j}(\omega_{0})\right)\rho_{nm}^{11}+ \notag\\
\frac{g^{2}}{\Omega^{2}}\left(G_{j}(-\omega_{+})\sqrt{n}\sqrt{m}\rho_{n-1,m-1}^{00}+G_{j}(-\omega_{-})\sqrt{n+1}\sqrt{m+1}\rho_{n+1,m+1}^{00}\right)+G_{j}(-\omega_{0})\rho_{n,m}^{00} \notag\\
\dot{\rho}_{nm}^{00}=\sum_{a}-\left(\frac{g^{2}}{2\nu^{2}}\left(G_{j}(-\omega_{+})(2+n+m)+G_{j}(-\omega_{-})(m+n)\right)+G_{j}(-\omega_{0})\right)\rho_{nm}^{00}+ \notag\\
\frac{g^{2}}{\nu^{2}}\left(G_{j}(\omega_{-})\sqrt{n}\sqrt{m}\rho_{n-1,m-1}^{11}+G_{j}(\omega_{+})\sqrt{n+1}\sqrt{m+1}\rho_{n+1,m+1}^{11}\right)+G_{j}(\omega_{0})\rho_{n,m}^{11}
\label{eq:supev}
\end{gather}

where the upper  indices	 of the density matrix refer to the TLS degrees
of freedom and the lower to the piston.

The bath-induced dynamics  associated with Eqs. \eqref{eq:supev} has two different time scales in the regime $\frac{g}{\nu}\ll1$:
the fast (change of the TLS states populations) and the slow (change of the
piston state). The ratio of populations in the excited
and ground  states quickly equilibrates for each element of the piston density
matrix $\rho_P$. This process is governed by:

\begin{gather}
\dot{\rho}_{nm}^{11}=-G(\omega_{0})\rho_{nm}^{11}+G(-\omega_{0})\rho_{nm}^{00}, \notag\\
\dot{\rho}_{nm}^{00}=-G(-\omega_{0})\rho_{nm}^{00}+G(\omega_{0})\rho_{nm}^{11}
\end{gather}

where $G(\omega)=\sum_{j=H,C}G_{j}(\omega)$ 

The partially steady
state TLS populations are

\begin{gather}
\tilde{\rho}^{11}=\sum_{n}\rho_{nn}^{11}=\frac{G(-\omega_{0})}{G(-\omega_{0})+G(-\omega)},
 \quad
\tilde{\rho}^{00}=\sum_{n}\rho_{nn}^{00}=\frac{G(\omega_{0})}{G(-\omega_{0})+G(-\omega)}; \notag\\
\rho_{nm}^{11}=\rho_{nm}\frac{G(-\omega_{0})}{G(-\omega_{0})+G(-\omega)}, \quad
\rho_{nm}^{00}=\rho_{nm}\frac{G(\omega_{0})}{G(-\omega_{0})+G(-\omega)}.
\label{eq:suss}
\end{gather}

The Fokker-Planck equation \eqref{M2} is isomorphic to the use of \eqref{eq:suss} in \eqref{eq:supev}, followed by the tracing-out of the TLS to obtain the evolution of P only.

\end{document}